\magnification=1200
\headline{\ifnum\pageno=1 \nopagenumbers 
\else \hss\number \pageno \fi} 
 
\overfullrule=0pt

\font\boldgreek=cmmib10 
\textfont9=\boldgreek
\mathchardef\mymu="0916 
\footline={\hfil}  
\baselineskip=10pt

\def\R{ {\rm R \kern -.31cm I \kern .15cm}}
\def\C{ {\rm C \kern -.15cm \vrule width.5pt \kern .12cm}}
\def\Z{ {\rm Z \kern -.27cm \angle \kern .02cm}}
\def\N{ {\rm N \kern -.26cm \vrule width.4pt \kern .10cm}}
\def\1{{\rm 1\mskip-4.5mu l} }
\def\tvi{\vrule height 12pt depth 6pt width 0pt}
\def\tv{\tvi\vrule}
\def \cc #1 {\kern .7em\hfill#1\hfill\kern .7em}
\vbox to 1,5truecm{}
\centerline{\bf Strangeness Enhancement Scenarios :}\medskip
\centerline{\bf Fireball or Independent Strings ?}
\vbox to 2 truecm {}
\centerline{\bf A. Capella}
\centerline{Laboratoire de Physique ThŽ\'eorique et Hautes Energies\footnote{*}{Laboratoire
associ\'eŽ au Centre National de la Recherche Scientifique - URA D0063}}
\centerline{Universit\'eŽ de Paris XI, b‰\^atiment 211, 91405 Orsay Cedex, France}  \medskip
 
\bigskip\bigskip
\baselineskip=20pt 
\noindent 
${\bf Abstract}$ \par 
Due to the long-standing discrepancy between NA35 and NA36 data on $\Lambda$ production,
two drastically different scenarios of strangeness enhancement are still possible.
Independent string models, such as the dual parton model, lead to results close to the
NA36 data. On the contrary, the NA35 results can only be described by introducing full
final state rescattering of the produced particles. The corresponding predictions for
central $Pb$-$Pb$ collisions at CERN energies differ by a factor 3 to 4. Preliminary
data on the net proton ($p$-$\bar{p}$) rapidity distribution in $Pb$-$Pb$ collisions
favor the independent string scenario. \par

\vbox to 2cm{}

\noindent LPTHE Orsay 96-30 \par 
\noindent April 1996
\vfill\supereject 
\vbox to 1 truecm {}
Most of the string model attempts to describe strangeness enhancement in heavy ion
collisions go beyond the strict framework of independent strings. The modifications
consist of string fusion [1-8] and, in most cases, final state rescattering of the
produced particles [3-8]. String fusion is quite natural in string models. In
particular, in the dual parton model [9] (DPM), string fusion is simulated by
$qq$-$\overline{qq}$ pairs from the nucleon sea [2,3], and, in this way, the string
independence is maintained. Final state rescattering, on the contrary, is a drastic
departure from independent strings models. When fully effective, it implies that the
results of these models can only be used as an initial condition - the final state being
strongly modified by the rescattering of the produced particles which drives the system
towards equilibrium. The model becomes then similar to the so-called fireball models -
such as hadron gas models and models based on the production of a deconfined quark-gluon
plasma [10-14]. \par

What do we learn from the confrontation of the string models with the available data~? A
first remark is that in the independent string framework (i.e. without final
state interaction), the effect of string fusion (or sea $qq$-$\overline{qq}$ pairs) is
numerically small when considering $\Lambda$ production. Indeed we know experimentally
that the ratio $\bar{\Lambda}/\Lambda$ at mid-rapidities is small in heavy ion
collisions. In central $SS$ collisions the most accurate value of this ratio is $0.24
\pm 0.01$ [15] and in $SW$ is $0.20 \pm 0.01$ [16]. We now know that in $Pb$-$Pb$ this
value is even smaller~: $0.154 \pm 0.005$ [17]. Since string fusion produces baryons in
pairs, it is clear that this mechanism can only affect the $\Lambda$ yield at 
10 $\%$ level. The situation is quite different in the presence of
final state rescattering. In this case $\Lambda$-$\bar{\Lambda}$ pair production from
string fusion can be much more important provided a sufficient number of $\bar{\Lambda}$'s
annihilate in collisions with nucleons. \par

Based mostly on NA35 results [18], it is widely accepted that a fireball scenario is
the only possible one - either with QGP formation or with full final state
rescattering. The only work where the fireball scenario has been explicitly dismissed
[8], uses, in order to describe the data, the VENUS Monte Carlo [7] which actually
contains final state rescattering. The HIJING Monte Carlo also used by the authors of
[8], which has no final state interaction, does not describe either the NA35 [18] or
the NA36 data [19] (see fig. 1c of ref. [8]). \par

In most analysis, the NA36 data on $\Lambda$ production are not
considered. This is due in part to their limited coverage in $p_{\bot}$. At present, NA36
data have been fully corrected for acceptance, efficiency and decay via unseen channels
[19]. (Corrected $\bar{\Lambda}$ and $K_0^s$ data are not given). Moreover, they
can be extrapolated to the full $p_{\bot}$ range, or, alternatively, the NA35 data can be
restricted to the $p_{\bot}$ range of the NA36 experiment so that a comparison of the two
data sets is possible. It turns out that there is a discrepancy between them which
exceeds a factor two [19]. \par

The purpose of this letter is to show that the NA36 data on $\Lambda$ production can
be described in the strict framework of DPM with independent strings. An important
ingredient in achieving this goal is the novel mechanism of baryon stopping
introduced in ref. [20] - which increases substantially the $\Lambda$ yield at
mid-rapidities. No final state rescattering is needed. When such rescattering is
introduced, the $\Lambda$ yield is increased by a factor 2~$\div$~2.5 and agreement with
the NA35 data is then achieved. It is also shown that the predictions of these two
scenarios for central $Pb$ $Pb$ collisions are dramatically different and thus a clear
experimental distinction will be possible. Moreover, recent preliminary data on
the net proton yield ($p$-$\bar{p}$) in central $Pb$-$Pb$ collisions [21] give some
indirect evidence in favor of the independent string scenario. Indeed, the DPM
prediction for this difference, with no final state interaction [20], is in good
agreement with these data. The final state interaction $\pi + N \to K +
\Lambda$, which is mainly responsible for the increase of the $\Lambda$ yield, produces a
corresponding decrease in the $N$ one which destroys the agreement between
theory and experiment. \par

We turn now to the calculation of the $\Lambda$ rapidity distribution in $pp$,
central $SS$ and central $Pb$-$Pb$ collisions at CERN energies. We proceed in the
framework of the dual parton model [3, 9]. In a first step we switch off the
$qq$-$\overline{qq}$ pairs in the nucleon sea ($\alpha = 0$). Moreover, we also
switch off all final state interactions. As discussed in [3] an important drawback
of the model is a too small baryon stopping. As a consequence, the rapidity
distributions of both proton and $\Lambda$ in central $SS$ collisions have a
pronounced dip at mid-rapidities not present in the data. This problem has been
solved in ref. [20] with the introduction of a diquark breaking (DB) component - in
which the baryon number follows one the valence quarks and is thus produced closer
to $y^* \simeq 0$. Based on ISR data on $p$ and $\bar{p}$ inclusive production
near $y^* \sim 0$, it was concluded [20, 22] that this DB component has a
cross-section $\sigma_{DB}^{pp} =$ 7 mb, and thus its size is about 20 $\%$ - the
diquark preserving (DP) component, corresponding to the fragmentation of the
diquark as a whole, being about 80~$\%$. An important result of ref. [20] is
that the DB component increases with $A$ faster than the ordinary (DP) one. As a
consequence, the effect of this mechanism on the proton yield in $NN$ and in
peripheral $SS$ collisions is rather small - while in central $SS$ collisions
it grows larger and has the right size needed to reach agreement with the data. The
prediction for central $Pb$-$Pb$ is also given in ref. [20]. Recent preliminary
data [21] nicely confirm both the plateau height and the shape of this
prediction. \par

With this novel mechanism of baryon stopping the formulae for the $\Lambda$
rapidity distribution in $pp$ and $AA$ collisions are
$${dN^{pp \to \Lambda} \over dy} (y) = {\sigma_{DB} \over \sigma_{in}} \
{dN_{DB}^{NN \to \Lambda} \over dy} (y) + \left ( 1 - {\sigma_{DB} \over \sigma_{in}}
\right ) {dN_{DP}^{NN \to \Lambda_L} \over dy}(y) + {dN_{DP}^{NN \to
\Lambda_{NL}} \over dy} (y) \eqno(1)$$

$${dN^{AA \to \Lambda} \over dy}(y) = \bar{n}_A^{DB} \left [ {dN_{DB}^{NN \to
\Lambda} \over dy} (y) \right ]_{n = \bar{n}/\bar{n}_A} + \left ( \bar{n}_A -
\bar{n}_A^{DB} \right ) \left [ {dN_{DP}^{NN \to \Lambda_L} \over dy}(y) \right
]_{n = \bar{n}/\bar{n}_A}$$ $$+ \bar{n}_A \left [ {dN_{DP}^{NN
\to \Lambda_{NL}} \over dy}(y) \right ]_{n = \bar{n}/\bar{n}_A} \ \ \ . \eqno(2)$$

\noindent Here $\sigma_{DB} =$ 7 mb, $\sigma_{in} =$ 32 mb, \par \nobreak
$\bar{n}_A^{DB} = A \int d^2b \left [ 1 - (1 - \sigma_{DB} \ T_{AA}(b))^A \right
]/\int d^2b \ \sigma_{AA}(b)\ ,$ \par \nobreak
$\bar{n}_A = A \int d^2b \left [ 1 - (1 - \sigma_{in} \ T_{AA}(b))^A \right
]/\int d^2b \ \sigma_{AA}(b) \ .$ \par 

\noindent The profile function is given by
$$T_{AA}(\vec{b}) = \int d^2s \ T_A(\vec{s}) \ T_A(\vec{b} - \vec{s}) \ \ \ .$$
\noindent In numerical calculations, a standard Saxon-Woods form has been used. For
central collision we take $b = 0$ (taking $b \leq$ 1 fm the differences are negligeably
small). $dN_{DP}^{\Lambda_L}/dy$ and $dN_{DP}^{\Lambda_{NL}}/dy$ are the ordinary (DP)
contributions for leading and non-leading $\Lambda$ respectively. The corresponding
formulae and numerical parameters are given in ref. [3]\footnote{*} {The non-leading
$\Lambda$ contribution is proportional to the number of participants, $\bar{n}_A$, and
not to the number of collisions, because, at CERN energies, the production of
$\Lambda$-$\bar{\Lambda}$ pairs in strings which do not involve a diquark is negligeably
small.}. Note that with $\sigma_{DB=0}$ one recovers the expressions of ref. [3]. The $DB$
component is [20] $$\left [ {dN_{DB}^{NN \to \Lambda} \over dy} (y) \right ]_n = C \left [
\widetilde{\rho}_{q_v}^{\ n} (y) + \widetilde{\rho}_{q_v}^{\ n}(-y) \right ] \ \ \ ,$$
$$\widetilde{\rho}_{q_v}^{\ n}(y) \equiv Z \ \rho_{q_v}^n(Z) = Z^{1/2} \int_Z^{1-Z} {dX
\over \sqrt{X}} (1 - X - Z)^{n - 3/2} \ \ \ . \eqno(3)$$ \noindent This contribution is
proportional to the momentum distribution of a valence quark in a proton - which in DPM
is entirely determined from reggeon intercepts [9, 20]. The constant $C$ is determined
from the normalization to 0.16. The origin of this value is the following. In a $NN$ or
$pp$ collision the $\Lambda$ average multiplicity is 0.1 [23]. However, in the $DB$
component, strangeness production is enhanced by a factor two, since the $s$-quark can be
produced on either side of the valence quark carrying the baryon quantum number [20].
Since the size of the $DB$ component in $pp$ is about 20 $\%$, we have to decrease by the
same amount the $\Lambda$ normalization in order to keep the average $\Lambda$
multiplicity in $pp$ collisions unchanged. This leads to the value $C = 0.8 \times 0.2 =
0.16$. Likewise the value of the normalization constant $a_{\Lambda}$ in the $\Lambda$
fragmentation function given in ref. [3] has to be reduced by 20 $\%$ - while the
corresponding values for $\bar{\Lambda}$, $p$ and $\bar{p}$ are unchanged. 
\par  

The results for $pp \to \Lambda$ are then very similar to the ones obtained
in fig. 1 of ref. [3] and agree with experiment both in shape and absolute normalization.
More precisely, we obtain $<n>_{\Lambda} = 0.11$ to be compared with the experimental
value $<n>_{\Lambda} = 0.096 \pm 0.01$ [23] and $(dN/dy)_{y^*=0} = 0.017$ to be compared
with $0.015 \pm 0.005$ [24]. \par

The numerical results for central $SS$ and central $Pb$-$Pb$ collisions are given in
the first columns of Tables~1 and 2. For central $SS$ collisions we see that the result is
close to the NA36 data. As for central $Pb$-$Pb$ collisions the prediction for the
plateau height is about 8 $\Lambda$'s per unity rapidity. \par

We turn next to the effect of the sea $qq$-$\overline{qq}$ pairs. The calculation of
this effect has large numerical uncertainties discussed in ref. [3]. The values given
there correspond to the maximal possible ones. As discussed above, in the absence of
final state interaction, the amount of $\Lambda$-$\bar{\Lambda}$ pair production due to
this mechanism is limited by the ratio $\bar{\Lambda}/\Lambda$. This value is small even
at $y^* \sim 0$ [15, 17]. The maximal number of $\Lambda$-$\bar{\Lambda}$ pairs computed
in ref. [3] would yield much larger values for this ratio. Renormalizing downward this
number of $\Lambda$-$\bar{\Lambda}$ pairs, in such a way
that the experimental ratio $\bar{\Lambda}/\Lambda$ is reproduced, we obtain the results
given in the second columns of Tables 1 and 2. As anticipated, the amount of pair
production both in $SS$ and $PbPb$ is very small. \par

We consider next the fireball or hadron gas scenario by introducing final state
interaction of the produced particles. In ref. [3] it was argued that the reactions
mainly responsible for the increase and decrease of the $\Lambda$ yield are $\pi + N
\to K + \Lambda$ and $\pi + \Lambda \to K + N$, respectively. Following [3] the excess
of $\Lambda$'s is given by 
$$\Delta \left [ {dN^{AA \to \Lambda} \over dy}(y) \right ] = \int d^2s {dN^{AA \to
\pi^-} \over dy \ d^2s} \left [ {dN^{AA \to p} \over dy \ d^2s} - {dN^{AA \to \Lambda}
\over dy \ d^2s} \right ] 3<\sigma> \ \ell n \left [ {\tau + \tau_0 \over \tau_0} \right
] \ . \eqno(4)$$
\noindent The particle densities in the rhs of (4) are those obtained in the model without
final state interaction (their explicit forms are given in [3]) and $<\sigma> =$ 1.5 mb
$\tau_0 =$ 1 fm and $\tau = 3$ fm. In the calculation one has to divide the $\ell n
\ \tau$ interval into a large number of subintervals (in practice a division into 10
equal subintervals gives a good accuracy). After each subinterval one has to evaluate the
new values of the $p$ and $\Lambda$ densities resulting from the final state
interaction. These values have to be used as initial conditions for the next
subinterval and so on. In order to do so, one has to know the
decrease in the proton yield associated to the increase, $\Delta$, in the $\Lambda$
one. If the only strange baryon produced by the final state interaction were
$\Lambda$'s, the decrease of the proton yield would be $\Delta/2$ - with a similar
decrease in the neutron one. However, since $\Sigma^{\pm}$, ... are also produced
(totalizing approximately the same excess $\Delta$), we assume that the decrease of
the proton yield is also equal to $\Delta$  [3]. The results for the $\Lambda$ rapidity
distribution after final state interactions are given in columns 4 and 5 of Table 1
for $SS$ and in columns 3 and 4 of Table 2 for $Pb$-$Pb$ central collisions. Again, the
first of these columns is the result without sea $qq$-$\overline{qq}$ pairs ($\alpha =
0$ in ref. [3]) and the second one is the result with sea $qq$-$\overline{qq}$ pairs as
computed in ref. [3] ($\alpha = 0.1$)\footnote{*}{As mentioned above, in the presence of
final state interaction, a large $\Lambda$-$\bar{\Lambda}$ pair production as the one
obtained in [3], can be consistent with the small experimental value of the ratio
$\bar{\Lambda}/\Lambda$ due to possible experimental annihilation with nucleons. Since
the calculation of [3] with ($\alpha = 0.1$) gives the maximal $\Lambda$-$\bar{\Lambda}$
production which is possible from sea $qq$-$\overline{qq}$ pairs, the number of
$\Lambda$'s has to be in between the values given in the two columns $\alpha = 0$ and
$\alpha = 0.1$.}. \par

In central $SS$ collision the $\Lambda$ yield has increased by a factor
2~$\div$~2.5 due to final state rescattering and is close to the NA35 data [18]. In
central $Pb$-$Pb$ collisions the final state interaction has produced a dramatic increase
in the $\Lambda$ yield - which is 3 to 4 times larger than the corresponding one without
final state interaction. In absolute value this difference ranges from 15 to 23 units.
Such a huge difference should be easy to detect experimentally. Data on the $\Lambda$
rapidity distribution in central $Pb$ $Pb$ collisions will soon be available. In the
meantime, it is important to note that there already exists some indirect evidence in
favor of the independent string scenario (i.e. DPM without final state interaction). As
already mentioned, preliminary results [21] on the net proton yield ($p$-$\bar{p}$) in
central $Pb$ $Pb$ collisions are in good agreement with DPM predictions without final
state interaction [20]. It has been shown above that the latter produces a substantial
increase in the $\Lambda$ yield and a corresponding decrease in the $N$ one (about 15
units in central $Pb$~$Pb$ at $y^* \sim 0$). The decrease in the number of $\bar{N}$ due
to $\pi + \bar{N} \to K + \bar{\Lambda}$ is considerably smaller. Therefore, as a
consequence of final state rescattering, the $p$-$\bar{p}$ yield will decrease destroying
the agreement between the DPM prediction and experiment. \par

In conclusion, it should be stressed that even if forthcoming CERN data confirm the DPM
prediction without final state interaction (i.e. a $\Lambda$ plateau height of about 8
units), production of fireballs or QGP droplets remains possible in events rearer than
the ones considered here. If this is the case it could affect the
production of (anti) cascades and (anti) omegas. However, the confirmation of the DPM
prediction with independent strings in central $Pb$ $Pb$ collisions at the level of
$\Lambda$ production would be quite striking and would confine the QGP search to a much
lower production level.  \vfill \supereject
\centerline{\bf Table 1} \bigskip
$$\vbox{\offinterlineskip \halign{
\tv#  &\cc{#} &\tv# &\cc{#} &\tv# &\cc{#} &\tv# &\cc{#} &\tv# &\cc{#} &\tv# &\cc{#} &\tv#
&\cc{#} &\tv# \cr \noalign{\hrule}
&$y^*$ && \multispan 3 $(dN/dy)^{SS \to \Lambda}_{no \ fsi}$
&&$(dN/dy)^{SS \to \Lambda}_{NA36}$ && \multispan 3 $(dN/dy)^{SS \to \Lambda}_{with
\ fsi}$ &&$(dN/dy)^{SS \to \Lambda}_{NA35}$  & \cr  \noalign{\hrule} &0$\ \ $  &&0.72 
&&0.78 &&0.97 $\pm$ 0.14 &&\ \ 1.5 &&1.9 &&2.2 $\pm$ 0.3 & \cr \noalign{\hrule}
&0.5  &&0.72 &&0.77 &&0.97 $\pm$ 0.12 &&\ \ 1.4 &&1.8 &&2.1 $\pm$ 0.3 & \cr
\noalign{\hrule}
&1$\ \ $  &&0.71 &&0.75 &&0.86 $\pm$ 0.10 &&\ \ 1.4 &&1.7 &&2.1 $\pm$ 0.3 & \cr
\noalign{\hrule}
&1.5  &&0.67 &&0.70 &&0.76 $\pm 0.12^*$ &&\ \ 1.4 &&1.6 &&2.2 $\pm$ 0.3 & \cr
\noalign{\hrule}
&2$\ \ $  && 0.61  &&0.62 && && \ \ 1.2  && 1.3 &&1.4 $\pm$ 0.2 & \cr
\noalign{\hrule}   
}}$$

$\Lambda$ rapidity distribution in central $SS$ collisions at 200 GeV/c per nucleon.
Columns 1 and 2 are the values without final state interactions - respectively without
and with sea $qq$-$\overline{qq}$ pairs (see main text). Columns 4 and 5 are the
corresponding values with final state interaction. The NA35 values are read from
Fig. 11b of [18] and those of NA36 from Fig.~14 of [19] with the $p_{\bot}$
acceptance correction given in eq. (1) (the value with an asterix is for $y^* = 1.25$).

\centerline{\bf Table 2} \bigskip
$$\vbox{\offinterlineskip \halign{
\tv#  &\cc{#} &\tv# &\cc{#} &\tv# &\cc{#} &\tv# &\cc{#} &\tv# &\cc{#} &\tv# \cr
\noalign{\hrule}
& $y^*$ && \multispan 3 $(dN/dy)^{Pb \ Pb \to \Lambda}_{no \ fsi}$
&& \multispan 3 $(dN/dy)^{Pb \ Pb \to \Lambda}_{with \ fsi}$ & \cr 
\noalign{\hrule}
& 0$\ \ $  &&\ \  7.7 && 8.4 && 23 && 31 &
\cr \noalign{\hrule}
& 0.5  &&\ \  7.4 && 8.1 && 22 && 30 & \cr
\noalign{\hrule}
& 1$\ \ $  &&\ \  6.5 && 6.8 && 20 && 25 & \cr
\noalign{\hrule}
& 1.5  &&\ \  4.9 && 5.0 && 16 && 19 & \cr
\noalign{\hrule}
& 2$\ \ $  && \ \ 2.9  && 2.9 && \ \ 8.8  && 9.1 & \cr
\noalign{\hrule}   
}}$$
 Same as Table 1 for central $Pb$ $Pb$ collisions at 160 GeV/c per nucleon.

 \vfill\supereject \centerline{\bf References} \bigskip 
\item{[1]} N. Armesto, M. A. Braun, E. G. Ferreiro
and C. Pajares, University of Santiago de Compostela, preprint US-FT/16-94. References to
earlier papers on string fusion can be found in G. Gustafson, Nucl. Phys. {\bf A566} (1994)
233c. 
\item{[2]} J. Ranft, A. Capella, J. Tran Thanh Van, Phys. Lett. {\bf B320} (1994)
346~; \item{} H. J. Mš\"ohring, J. Ranft, A. Capella, J. Tran Thanh Van, Phys. Rev.
{\bf D47} (1993) 4146 (the calculations in these papers are based on the DPMJET and
DTNUC codes).  
\item{[3]} A. Capella, A. Kaidalov, A. Kouider Akil, C. Merino and J. Tran Thanh Van, Z.
Physik C, in press.
\item{} A. Capella, Phys. Lett. {\bf B364} (1995) 175.
\item{} A. Capella, A. Kaidalov, A. Kouider Akil, C. Merino, J. Ranft and J. Tran Thanh
Van, Proceedings XXX Rencontres de Moriond, Les Arcs, France (1995).
  \item{[4]} QGSM~: A. Kaidalov,
Phys. Lett. {\bf B117} (1982) 459~; A. Kaidalov and K. A. Ter-Martirosyan, Phys. Lett.
{\bf B117} (1982) 247. For the corresponding Monte Carlo code see N. S. Amelin et al,
Phys. Rev. {\bf C47} (1993) 2299~; N. S. Amelin et al, Nucl. Phys. {\bf A544} (1992)
463c.     \item{[5]} RQMD~: H. Sorge, R. Matiello, A. von Kectz, H. Stšcker and W.
Greiner, Z. Phys. {\bf C47} (1990) 629~; H. Sorge, M. Berenguer, H. Stšcker and W.
Greiner, Phys. Lett. {\bf B289} (1992) 6~; Th. Schšnfeld et al, Nucl. Phys. {\bf A544}
(1992) 439c~; H. Sorge, Z. Phys. {\bf C67} (1995) 479.     
\item{[6]} FRITIOF~: B. Andersson, G. Gustafson and B. Nilsson-Almquist, Nucl . Phys. {\bf
B 281} (1987) 289~; B. Nilsson-Almquist, E. Stenlund, Comp. Phys. Comm. {\bf 43} (1987)
387. In this model the enhancement of $\Lambda$ and $\bar{\Lambda}$ is essentially due to
the final state in\-te\-raction~: A. Tai, Bo Andersson and Ben-Hao Sa, Proceedings of the
Strangeness '95 International Conference, Tucson, Arizona (1995).   
\vfill \supereject
\item{[7]} VENUS~: K.
Werner, Phys. Rep. {\bf 232} (1993) 87. \nobreak \item{} K. Werner and J. Aichelin, Phys.
Rev. Lett. {\bf 76} (1996) 1027.
  \item{[8]} V. Topor Pop et al, Phys. Rev. {\bf C52} (1995)
1618. Comments on this paper can be found in M. Ga\'zdzicki and U. Heinz preprint
IKF-HENPG/1-96.
  \item{[9]} A.
Capella, U. Sukhatme, C. I. Tan and J. Tran Thanh Van, Phys. Rep. {\bf 236} (1994) 225.  
\item{[10]} B. Koch, B. Muller and J. Rafelski, Phys. Rep. {\bf 142} (1986) 167. 
\item{[11]} J. Cleymans, K. Redlich, H. Satz and E. Suhonen, Z. Phys. {\bf C58} (1993)
347~;
\item{} J. Cleymans, D. Elliot, H. Satz and R. L. Thews, CERN TH 95-298.
 \item{[12]} J. Lettessier, A. Tounsi, U. Heinz, J. Sollfrank and J. Rafelski, Phys. Rev.
{\bf D51} (1995) 3408~; 
\item{} J. Sollfrank, M. Ga\'zdzicki, U. Heinz and J. Rafelski, Z. Phys. {\bf C61}
(1994) 659.
\item{} J. Rafelski et al, AIP Proceedings series V330 (1995) 490.
 \item{[13]} M. Ga\'zdzicki and D. Ršhrich, preprint IKF-HENPG/8-95.
\item{[14]} U. Ornick, M. PlŸmer, B. R. Schlei, D. Strottman, and R. M. Weiner, GSI
preprint 95-62.
 \item{[15]} S. Abatzis et al (WA94 coll), Phys. Lett. {\bf 354} (1995) 178.
\item{[16]} S. Abatzis et al (WA 85 coll), Phys. Lett. {\bf B359} (1995) 382.
\item{[17]} M. A. Mazzoni et al (WA 97 coll), Proceedings XXV International Symposium on
Multiparticle Dynamics, Star\'a Lesn\'a, Slovakia (1995). 
\item{} R. Lietava (WA97 coll), Proceedings XXXI Rencontres de Moriond, Les Arcs, France
(1996). \item{[18]} T. Alber et al (NA 35 coll), Z. Phys. {\bf C64} (1994) 195.
\item{[19]} E. G. Judd (NA36 coll), Nucl. Phys. A {\bf 590} (1995) 291c.
\item{[20]} A. Capella and B. Z. Kopeliovich, preprint LPTHE 96-01, hep-ph 9603279.
\item{[21]} P. Seyboth (NA49 coll), Proceedings XXXI Rencontres de Moriond, Les Arcs,
France (1996). \item{[22]} B. Z. Kopeliovich and B. G. Zakharov, Z. Phys. {\bf C43}
(1989) 241. \item{[23]} M. Ga\'zdzicki and Ole Hansen, Nucl. Phys. {\bf A528} (1991) 754.
\item{[24]} K. Jaeger et al, Phys. Rev. {\bf D11} (1975) 2405.
 \bye